\documentclass[10pt,aps,showpacs,floatfix,raggedbottom,floats,superscriptaddress,twocolumn]{revtex4-1}
\pdfoutput=1
\usepackage{graphicx,latexsym}
\usepackage{dcolumn}
\usepackage{amssymb,amsmath,bm}

\usepackage[hidelinks=true,breaklinks=true]{hyperref}
\usepackage{natbib}
\hypersetup{
  colorlinks   = true, 
  urlcolor     = blue, 
  linkcolor    = blue, 
  citecolor    = red 
}

\newcommand\mytop[2]{\genfrac{}{}{0pt}{}{#1}{#2}}
\DeclareMathOperator{\vctrz}{vec}

\begin{document}

\title{Cavity-photon induced high order transitions between ground states of quantum dots}

\author{Vidar Gudmundsson}
\email{vidar@hi.is}
\affiliation{Science Institute, University of Iceland, Dunhaga 3, IS-107 Reykjavik, Iceland}
\author{Nzar Rauf Abdullah}
\affiliation{Physics Department, College of Science, 
             University of Sulaimani, Kurdistan Region, Iraq}
\affiliation{Komar Research Center, Komar University of Science and Technology, Sulaimani, Kurdistan Region, Iraq}
\author{Chi-Shung Tang}
\email{cstang@nuu.edu.tw}
\affiliation{Department of Mechanical Engineering, National United University, Miaoli 36003, Taiwan}
\author{Andrei Manolescu}
\email{manoles@ru.is}
\affiliation{School of Science and Engineering, Reykjavik University, Menntavegur 
             1, IS-101 Reykjavik, Iceland}
\author{Valeriu Moldoveanu}
\email{valim@infim.ro}
\affiliation{National Institute of Materials Physics, PO Box MG-7, Bucharest-Magurele, Romania}

%

\begin{abstract}
We show that quantum electromagnetic transitions to high orders are essential to 
describe the time-dependent path of a nanoscale electron system in a Coulomb blockage
regime when coupled to external leads and placed in a three-dimensional rectangular photon 
cavity. The electronic system consists of two quantum dots embedded asymmetrically in a
short quantum wire. The two lowest in energy spin degenerate electron states are mostly localized 
in each dot with only a tiny probability in the other dot. In the presence of the leads we identify 
a slow high order transition between the ground states of the two quantum dots. 
The Fourier power spectrum for photon-photon correlations in the steady state shows a
Fano-type of a resonance for the frequency of the slow transition.
Full account is taken of the geometry of the multi-level electronic system,
and the electron-electron Coulomb interactions together with the para- and diamagnetic 
electron-photon interactions are treated with step wise exact numerical diagonalization and
truncation of appropriate many-body Fock spaces. The matrix elements for all interactions are
computed analytically or numerically exactly.
\end{abstract}

\maketitle
%
%

\section{Introduction}
Many research groups are studying the physics of strong and ultrastrong coupling
of light and matter in atomic or circuit-QED systems, or semiconductor heterostructures with the aid of photon 
cavities \cite{Peterson380:2012,Niemczyk10:772,ates2009:724,Yoshie2004,Reithmaier2004}. 
Some aspects of this research has been addressed in review articles recently 
\cite{0953-8984-29-43-433002,Frisk-Kockum2019}.
The research is carried out with multiple aim or goals in mind, ranging from optoelectronic 
or quantum computing devices to a convenient platform to study fundamental aspects of
strong matter-light interactions.

In a recent article Zhang et al.\ \cite{Zhang1005:2016} stress the importance of the possibilities of the
two-dimensional electron system (2DES) in the conduction band of a GaAs heterostructure to obtain  
a collective nonperturbative coupling of 2D electrons with high-quality-factor terahertz cavity photons.
They report experiments using cyclotron resonances between Landau levels to achieve both a small 
cavity-photon decay constant, and a low electron decoherence rate. Interestingly, they find that
a model including the diamagnetic electron-photon interaction best fits their results.
The electron system in the cavity is not
coupled to external leads functioning as electron reservoirs. Theoretically \cite{Ciuti05:115303}, 
and experimentally \cite{PhysRevLett.90.116401}, intersubband transitions in cavities have been
studied earlier.

The addition of metallic electron reservoirs has been experimented with \cite{0953-8984-29-43-433002},
which could lead the way to hybrid electron-photonic transport systems, that would still enhance the 
possible utility of the systems in devices and fundamental research.

In the last five years we have been developing an approach to model the time-dependent
transport of electrons through a nanoscale electronic system embedded in a photon cavity.
The main emphasis has been on multilevel systems with a specific shape or geometry,
higher order interactions between electrons, and electrons and photons, and a time scale
spanning the transient to the steady-state regime. To accomplish this we have used a
generalized master equations and exact numerical diagonalization to study various
effects \cite{Nzar-2019-Rabi,GUDMUNDSSON20181672,2017arXiv170603483G,PhysRevB.97.195442}.

At the formal level, we refine the calculation of both the para- and the diamagnetic 
contributions to the electron-photon interaction. The corresponding matrix elements are calculated 
exactly for a specific type of a cavity, a three-dimensional rectangular one. By doing so we are able 
to test the validity of the usual lowest order approximation with respect to the ratio of the size 
of the 3D cavity to the length of the central electronic system. This is an important step, especially 
in order to establish the correct form of the diamagnetic (i.e.\ A-square) term.
Second, and more important, we want to stress that nonperturbative approach to the 
electron-photon interactions is not only necessary for the strong and the ultrastrong
coupling regime, but also when electromagnetic transitions or tunneling through photon active states 
can take a very long time. This can be the case for transitions in a terahertz cavity containing
a GaAs heterostructure with active intersubband processes. 

Our central system has two well separated quantum dots so that 
we find the two states with the lowest energy (each one with their two spin components)
almost entirely localized in either dot, with only a tiny overlap with the other.
We will show that when the dot lower in energy is in a Rabi-resonance 
with the excited states within the bias window set by the external leads 
there is still a very slow high order photon active transition between the ``ground states'' of the dots.

The model and the improved electron-photon coupling are introduced in Section \ref{Sec2}, 
the transport formalism along with the discussion of the numerical results are presented in Section \ref{Sec3}. 
We conclude in Section \ref{Sec4} and include for completeness some technical results in two Appendices.

\section{The closed central system}
\label{Sec2}
We use a potential to describe the short parabolic quantum wire with two asymmetrically embedded
quantum dots \cite{Gudmundsson19:10}
\begin{align}\label{Potential}
      V(x,y) =& \left[\frac{1}{2}m^*\Omega_0^2y^2 -eV_g + \theta\left(\frac{L_x}{2}-|x|\right)\right. \\
      \times& \left. \sum_{i=1}^2 V_d^i\exp{\left\{-\beta_i^2 (x-x_{0i})^2-\beta_i^2(y-y_{0i})^2\right\}} 
      \right]\nonumber
\end{align}
with the parameters $\hbar\Omega_0 = 2.0$ meV, $V_d^1 = -6.6$ meV, $V_d^2 = -6.8$ meV,
$\beta_1 = 0.030$ nm$^{-1}$, $\beta_2 = 0.028$ nm$^{-1}$, 
$x_{01}=-48$ nm, $x_{02}=+48$ nm, $y_{01}=-50$ nm, $y_{02}=+50$ nm,
$L_x = 180$ nm. $\theta$ is the Heaviside unit step function, and $V_g$ is the plunger gate voltage
used to shift the energy of electronic states with respect to the bias window that will be defined
by the chemical potentials of the left (L) and right (R) leads.

In terms of field operators for the electrons in the conduction band of GaAs the 
current density and the probability density operators are
\begin{equation}
      \mathbf{j} = -\frac{e}{2m}\left\{\psi^\dagger\bm{\pi}\psi 
                 + \bm{\pi}^*\psi^\dagger\psi\right\}, \quad
      \rho = \psi^\dagger\psi ,           
\end{equation}
with 
\begin{equation}
      {\bm{\pi}}=\left(\mathbf{p}+\frac{e}{c}\mathbf{A}_{\mathrm{ext}}\right),
\label{pAext}
\end{equation}
where the external homogeneous magnetic field perpendicular to the plane of the central
two-dimensional (2D) system is represented by the classical vector potential 
$\mathbf{A}_{\mathrm{ext}}=(-By,0,0)$. The external magnetic field $B=0.1$ T and the 
parabolic confinement energy $\hbar\Omega_0=2.0$ meV in the $y$-direction for the 
central system (and the semi-infinite external leads) define a convenient length
scale $a_w=(\hbar /(m^*\Omega_w))^{1/2}$, where $\hbar\Omega_w=\hbar({\omega_c^2+\Omega_0^2})^{1/2}$,
and $\omega_c=(eB_{\mathrm{ext}})/(m^*c)$. For the GaAs parameters used here, 
$m^*=0.067m_e$, $\kappa_e=12.4$, and $g^*=-0.44$, we have $a_w\approx 23.8$ nm.
The main role of the external magnetic field is to lift spin and orbital degeneracies
without introducing considerable orbital magnetic effects for the small system.
The Hamiltonian for the central system is
\begin{align}\label{HS}
      H_\mathrm{S} =& \int d^2r \psi^\dagger (\mathbf{r})\left\{\frac{\pi^2}{2m^*}+
        V(\mathbf{r})\right\}\psi (\mathbf{r}) \nonumber\\
        +&H_\mathrm{EM} + H_\mathrm{Coul}+H_\mathrm{Z}\nonumber\\ 
        +&\frac{1}{c}\int d^2r\;\mathbf{j}(\mathbf{r})\cdot\mathbf{A}_\gamma
        +\frac{e^2}{2m^*c^2}\int d^2r\;\rho(\mathbf{r}) A_\gamma^2,
\end{align}
where $H_\mathrm{EM}=\hbar\omega a^\dagger a$ is the Hamiltonian for the single cavity mode
with energy $\hbar\omega$, $H_\mathrm{Z}$ is the Zeeman term for the electrons, and $H_\mathrm{Coul}$ is the 
mutual Coulomb interaction of the electrons with kernel
\begin{equation}
      V_{\mathrm{Coul}}(\mathbf{r}-\mathbf{r}') = \frac{e^2}{\kappa_e\sqrt{|\mathbf{r}-\mathbf{r}'|^2+\eta_c^2}},
\label{VCoul}
\end{equation}
and a small regularization parameter $\eta_c/a_w=3\times 10^{-7}$.

The first term in the third line of Eq.\ (\ref{HS}) is the paramagnetic electron-photon
interaction, while the second term is the diamagnetic part of the interaction.
\begin{figure}[htb]
      \centerline{\includegraphics[width=0.48\textwidth]{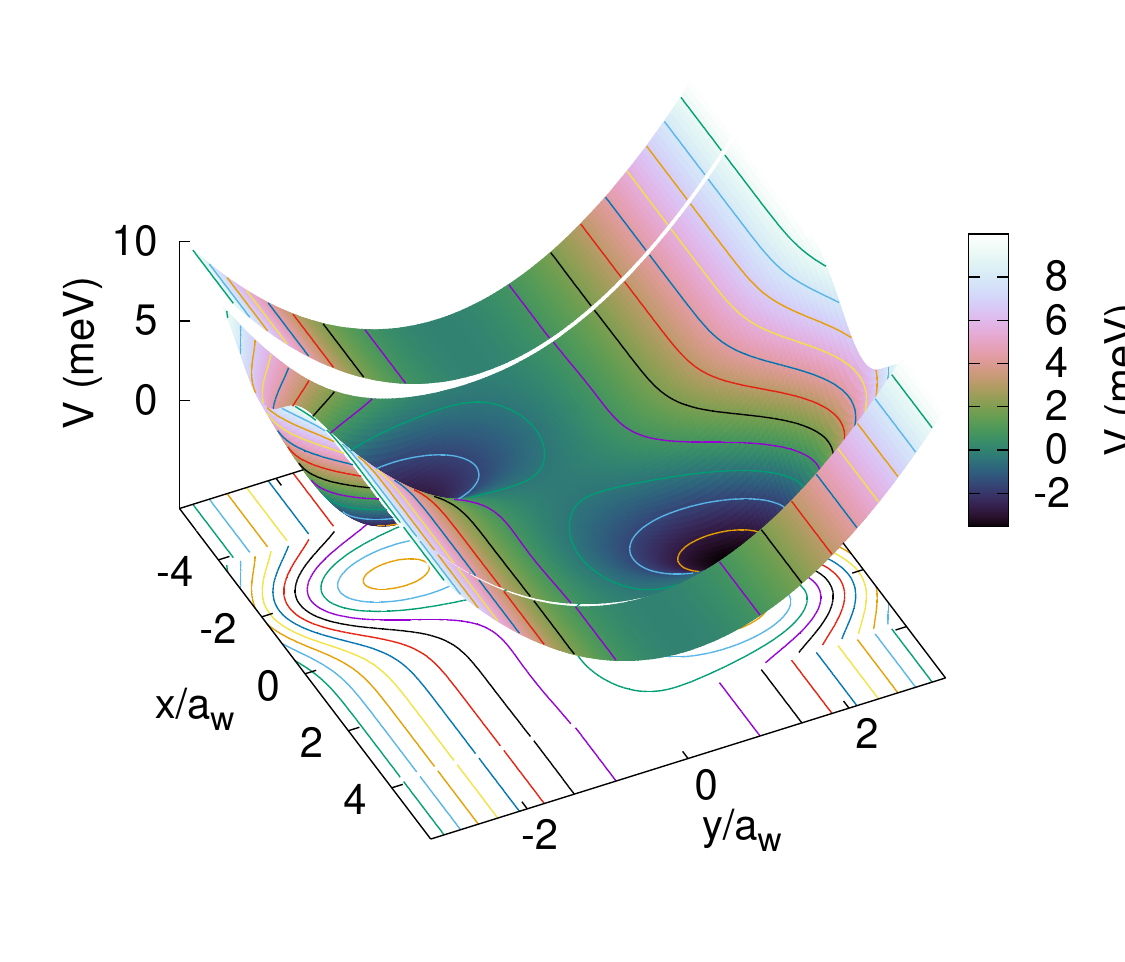}}
      \caption{The potential defining the short parabolic quantum wire with 
               two asymmetrically embedded quantum dots. The length of the
               short wire is $L_x=180$ nm. $a_w=23.8$ nm, and the short wire extends
               from $x/a_w\approx -3.8$ to $x/a_w\approx +3.8$, beyond which limits are indicated 
               the semi-infinite leads (separated by narrow gaps in the figure) that are coupled 
               to the short quantum wire at $t=0$.}
      \label{Fig01}
\end{figure}

We assume a rectangular photon-cavity of dimensions $a_\mathrm{c}\times b_\mathrm{c}\times d_\mathrm{c}$,
and use a Coulomb gauge for the quantized vector potential $\mathbf{A}_\gamma$ of the single-mode photon
field of the cavity. The polarization of the electric field of the cavity photons parallel to the 
transport in the $x$-direction (with the unit vector $\mathbf{e}_x$) can be realized in the TE$_{011}$ mode, 
or perpendicular to it (defined by the unit vector $\mathbf{e}_y$) by the TE$_{101}$ mode. 
The quantized vector potential for the two polarizations for the cavity field can be expressed 
(in a stacked notation) as \cite{Gudmundsson19:10}
\begin{equation}
      \mathbf{A}_\gamma (\mathbf{r})=\left(\mytop{\hat{\mathbf{e}}_x }{ \hat{\mathbf{e}}_y}\right)
      {\cal A}\left\{a+a^{\dagger}\right\}
      \left(\mytop{\cos{\left(\frac{\pi y}{b_\mathrm{c}}\right)}}{ 
      \cos{\left(\frac{\pi x}{a_\mathrm{c}}\right)}} \right)
      \cos{\left(\frac{\pi z}{d_\mathrm{c}}\right)},
\label{Cav-A}
\end{equation}
where the strength of the vector potential, ${\cal A}$,
and the electron-photon coupling constant are related by $g_\mathrm{EM} = e{\cal A}\Omega_wa_w/c$.
With the vector potential (\ref{Cav-A}) and the fermionic field operators for the electrons
the Hamiltonian for the electron-photon interactions becomes 
\begin{align}
\label{H-e-EM-q}
      H_\mathrm{e-EM}=&g_\mathrm{EM}\sum_{ij}d^\dagger_id_j g^p_{ij}\{a+a^\dagger \}\\ \nonumber
      +&\frac{g_\mathrm{EM}^2}{\hbar\Omega_w}\sum_{ij}d^\dagger_id_j g^d_{ij}
      \left[\left(a^\dagger a+\frac{1}{2}\right) +\frac{1}{2}(a^\dagger a^\dagger +aa)\right] .
\end{align}
When the approximation that the wavelength of the cavity field is much larger than the size
of the electronic system the second term of Eq.\ (\ref{H-e-EM-q}) becomes diagonal in the
electronic creation and annihilation operators, i.e.\ $g^d_{ij}\rightarrow\delta_{i,j}$
\cite{Jonasson2011:01,PhysRevE.86.046701,Gudmundsson12:1109.4728}. 
(It is important to note that the total diamagnetic interaction term is not diagonal,
and thus we have found that it can lead to a very weak Rabi-splitting when the paramagnetic
term is blocked by symmetry \cite{doi:10.1002/andp.201700334})
We will not make this approximation here, but use the fact that the original functional basis used
to construct the single-electron basis for the short parabolic quantum wire \cite{PhysRevE.86.046701}
can be used to obtain the exact matrix elements for the electron-photon interaction analytically
in a closed form. Technical details for the matrix elements, the energy spectra, and the many-body
states are found in Appendix \ref{AppA}.

\begin{figure}[htb]
	\centerline{\includegraphics[width=0.48\textwidth]{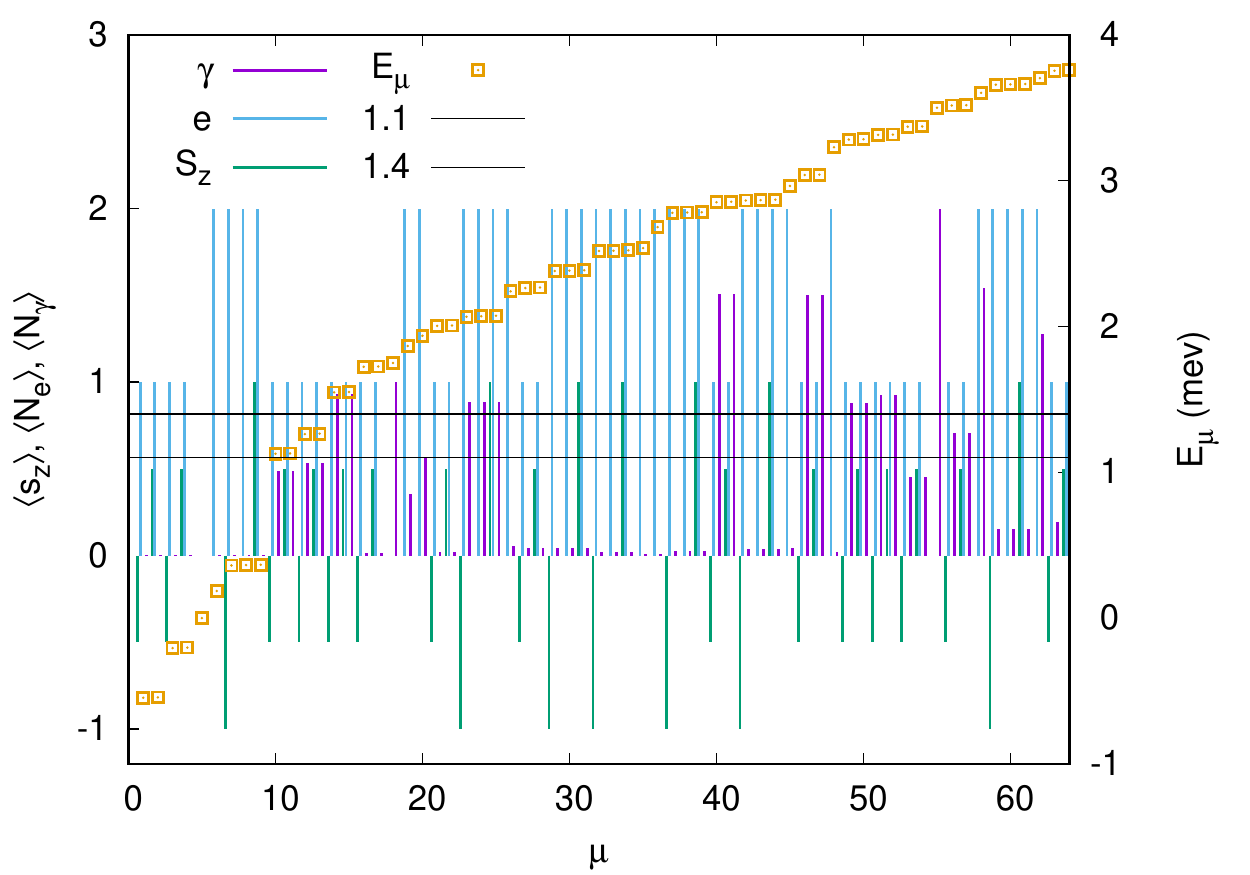}}
	\caption{The energy (squares, right axis), the mean electron number (e), the mean photon number
	         ($\gamma$), and the $z$-component of the spin, $S_z/\hbar$, as functions of the number of the many-body state,
             $\mu$. The horizontal black lines indicate the chemical potentials of the left, $\mu_L=1.4$ meV, and 
             right leads, $\mu_R=1.1$ meV. $g_\mathrm{EM} =0.05$ meV, $B=0.1$ T, $L_x=180$ nm,
             $-eV_g=0.2$ meV, and $\hbar\omega =1.75$ meV.}
	\label{Fig02}
\end{figure}

In order to understand better the structure or properties of the states $|\breve{\mu})$ for one value
of the coupling constant ($g_\mathrm{EM} =0.05$ meV) we show in Fig.\ \ref{Fig02} the energy of the 
lowest 64 states, together with their electron content, their spin $z$-component, and their mean photon 
number.
We note that the lowest 4 states are one-electron states of which the lower ones with opposite spin
$z$-component are mostly confined to the right quantum dot, while the higher ones are mostly confined
to the left dot \cite{Gudmundsson19:10}. The fifth state is the vacuum or the empty state, while the 
sixth state is the spin-singlet two-electron ground state and $|\breve{7})$, $|\breve{8})$, and $|\breve{9})$
are the lowest in energy spin-triplet two-electron states. The next four states, $|\breve{10})$, $|\breve{11})$, 
$|\breve{12})$, and $|\breve{13})$ are the first replica of the one-electron ground state and the
first excitation there of very close to a Rabi resonance, as can be verified by their mean photon
content that is close to 1/2. These four states are in the bias window defined by the chemical potentials
of the left and right leads. The next two states, $|\breve{14})$ $|\breve{15})$, are the first photon 
replicas of the one-electron states $|\breve{3})$ $|\breve{4})$ mostly localized in the left dot.
The first excitations of $|\breve{3})$ $|\breve{4})$ are the next two states $|\breve{16})$ $|\breve{17})$
and we notice that the photon content of these four last states is close to an integer indicating
that they are not close to a Rabi-resonance, but they are very weakly coupled by a Rabi resonance.

The position of the bias window gives a hint that the system most likely will approach a Coulomb
blockage in the long time limit after the central system will be coupled to the external leads.
Exactly, how this happens and the very different role of the four one-electron states below the 
bias window in the time evolution of the coupled system will be the subject of the study presented
below.

\begin{figure}[htb]
	\centerline{\includegraphics[width=0.48\textwidth]{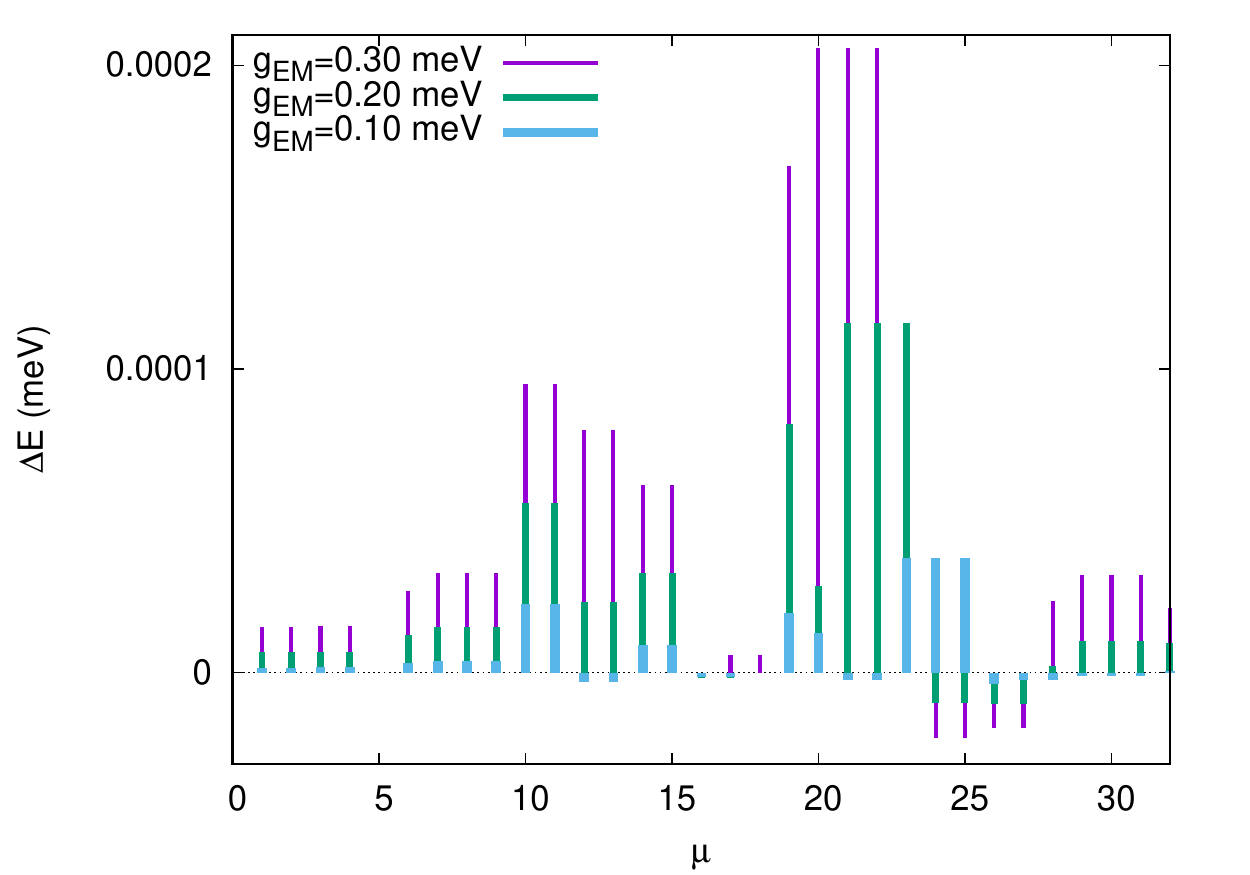}}
	\centerline{\includegraphics[width=0.48\textwidth]{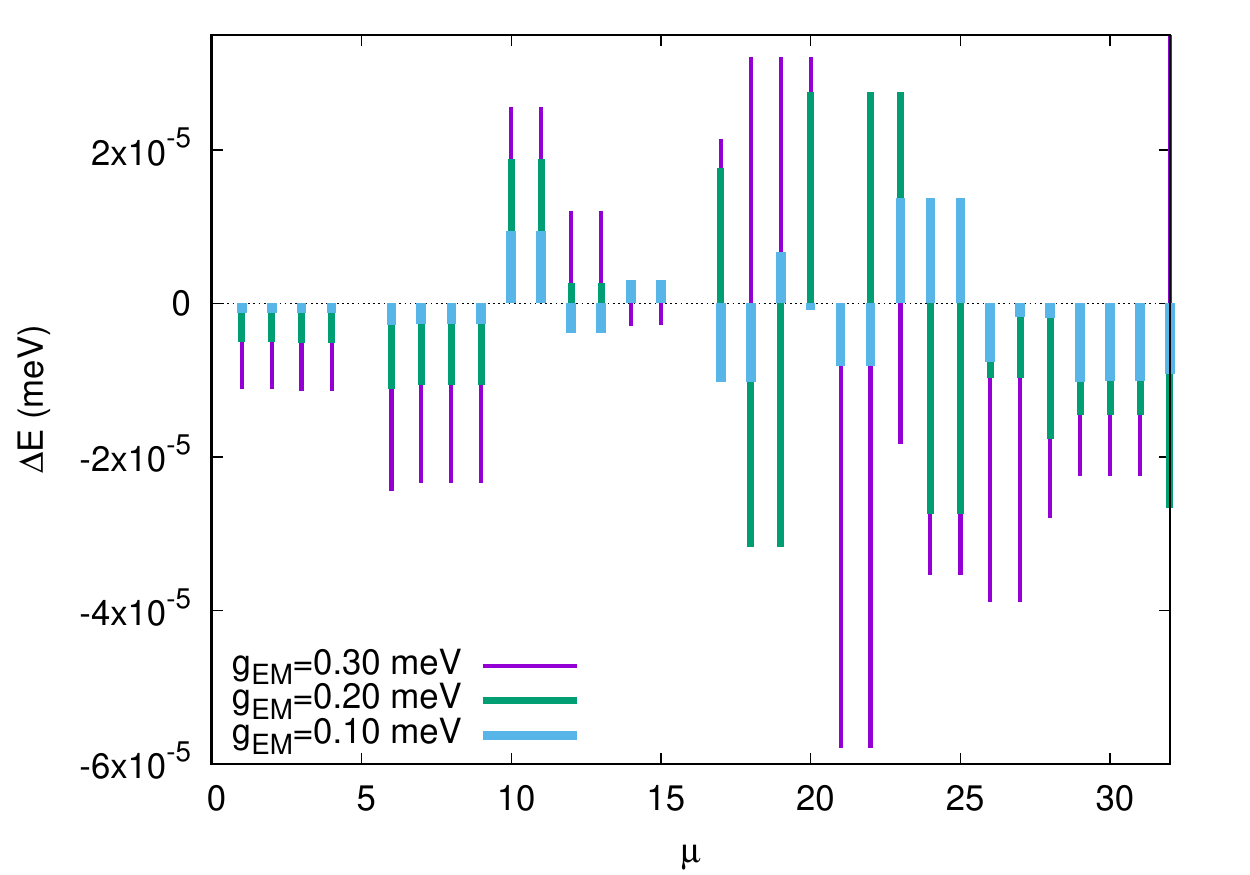}}
	\caption{The change in self-energy caused by the exact form of the paramagnetic and diamagnetic 
	         electron-photon interactions 
	       	 of each many-body state $|\breve{\mu})$ for $x$- (upper) and $y$-polarized (lower) cavity
	       	 field for three values of the coupling coefficient $g_\mathrm{EM}$. $B=0.1$ T, $L_x=180$ nm, 
	       	 $\hbar\Omega_0=2.0$ meV, $-eV_g=0.2$ meV, and $\hbar\omega =1.75$ meV.}
	\label{Fig03}
\end{figure}

For the cavity with photon energy $\hbar\omega =1.75$ meV the ratio of the size of the 
cavity to the length of the central system, $a_c/L_x$ or $b_c/L_x$, is 48.4 suggesting that
corrections to the matrix elements of the electron-photon interaction due to the variation
of the fields within the system size are minimal. This can indeed be verified by inspecting
the change in self-energies of the lowest 32 many-body states presented in Fig.\ \ref{Fig03}.
This last statement should though be enjoyed with care. When the fields are assumed constant within
the central electronic system the diamagnetic electron-photon interaction is diagonal and mainly 
contributes to the self-energy of the states, now with the exact matrix elements the contribution 
of the paramagnetic interaction to the self-energies is increased considerably, and the diamagnetic
interaction is not anymore diagonal. These small changes in the form of the interactions can be of
importance when investigating the very long time evolution of the system towards a steady state.
An other important point is that the rather artificial diagonal form of the diamagnetic interaction
is replaced. Results with the exact diamagnetic interaction can thus be used to verify how appropriate
that form is. Below, we will use results from the time evolution of the system to learn to appreciate
how important higher order electromagnetic transitions are to understand its properties and how far we are
forced away from a perturbational view of the underlying processes. 

Not surprisingly, we observe in Fig.\ \ref{Fig03} that the change in self-energy of the many-body states
of the central system is larger for the $x$-polarized photon field. This reflects the anisotropy of
the system, which is more easily polarized in the $x$-direction, that will be the direction of
transport through it. Furthermore, we note that the change in the self-energy is nonlinear with 
$g_\mathrm{EM}$, even though $g_\mathrm{EM}$ is not large compared to the confinement energy
$\hbar\Omega_0$.

\section{The electron transport, model and results}
\label{Sec3}
The central system is opened up for electron transport through it by coupling the left and right
semi-infinite quasi-one-dimensional leads to it at $t=0$. The leads are assumed to have the same
parabolic confinement in the $y$-direction as the central system and are subject to the same external
homogeneous magnetic field. The coupling to the leads is described by the Hamiltonian
\begin{equation}
    H_T = \theta (t)\sum_{il} \int d  \mathbf{q} \left(T_{\mathbf{q}i}^l c_{\mathbf{q}l}^\dagger d_i +
          (T_{\mathbf{q}i}^l )^* d_i^\dagger c_{\mathbf{q}l}\right),
\label{H_T}
\end{equation} 
where electrons in the short wire are created and annihilated by the operators $d_i^\dagger$ and $d_i$,
respectively, but in the leads by the operators $c_{\mathbf{q}l}^\dagger$ and $c_{\mathbf{q}l}$.
The quantum number $\mathbf{q}$ represents both the continuous momenta in the leads and a subband
index. The coupling tensor $T_{\mathbf{q}i}^l$ is calculated using the probability density of
each single-electron state of the lead $l$ and the central electron system in the contact region 
that is defined to extend approximately one $a_w$ into each subsystem 
\cite{Gudmundsson09:113007,Moldoveanu09:073019,Gudmundsson12:1109.4728}. 
At the same time, $t=0$, a photon reservoir of zero temperature is weakly coupled to the cavity
with coupling coefficient $\kappa$. As the temperature of the photon reservoir is zero
we can regard $\kappa$ as a cavity decay constant. The technical details for the method used
to derive the master equation used to describe the transport calculations are found in Appendix \ref{AppB}.

\begin{figure}[htb]
	\centerline{\includegraphics[width=0.40\textwidth]{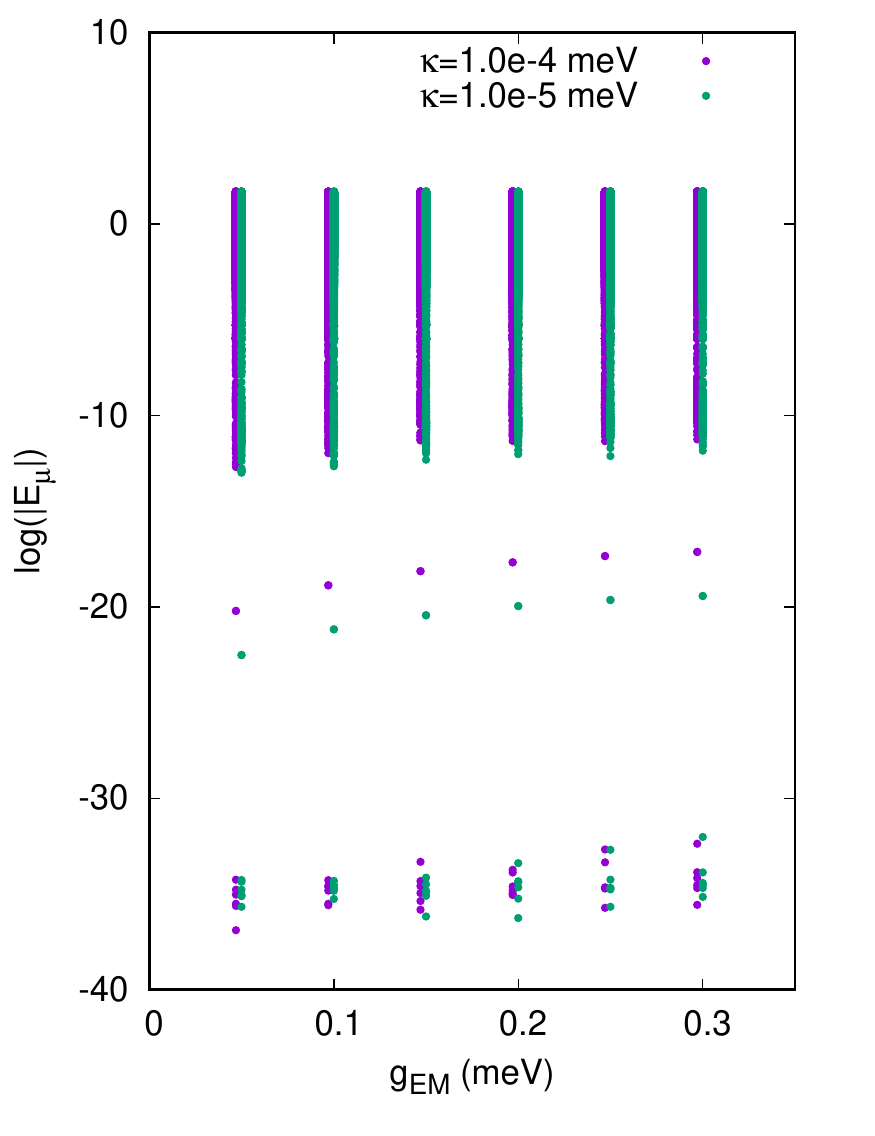}}
	\caption{The logarithm of the absolute values of the complex eigenspectrum of the 
	         Liouville operator $\mathfrak{L}$ as a function of the electron-photon coupling coefficient
             $g_\mathrm{EM}$ for two values of the cavity-environment coupling $\kappa$. 
             $B=0.1$ T, $-eV_g=0.2$ meV, and $L_x=180$ nm.}
	\label{Fig04}
\end{figure}

The structure of the master equation (\ref{finalME})
associates a zero eigenvalue with the steady state of the system \cite{Petrosky01032010,Nakano2010}. 
We thus display in Fig.\ \ref{Fig04} the logarithm of the absolute value of the eigenspectrum of 
$\mathfrak{L}$ as a function of the coupling parameter $g_\mathrm{EM}$.

If we fix the gate voltage such that both spin components of the one-electron ground state
are inside the bias window the system will not enter a Coulomb blockage regime, but conduct
at a constant average rate in the steady state. In this case we always find a single zero
eigenvalue. Here, when the system enters a Coulomb blockage regime we find a small null space
of $\mathfrak{L}$ well separated by a spectral gap. Within the spectral gap we find one eigenvalue
representing a very slow final transition in the system, that we will identify below.
No transition in the null-space can be ignored as that would lead to a violation of the condition
that $\mathrm{Tr_\mathrm{S}\{\rho_\mathrm{S}}\}=1$. The eigenvalues within the null-space are all
zero within the machine and software accuracy that can be expected, even though they show some spread on the 
logarithmic scale adopted in Fig.\ \ref{Fig04}. 

\begin{figure}[htb]
	\centerline{\includegraphics[width=0.48\textwidth]{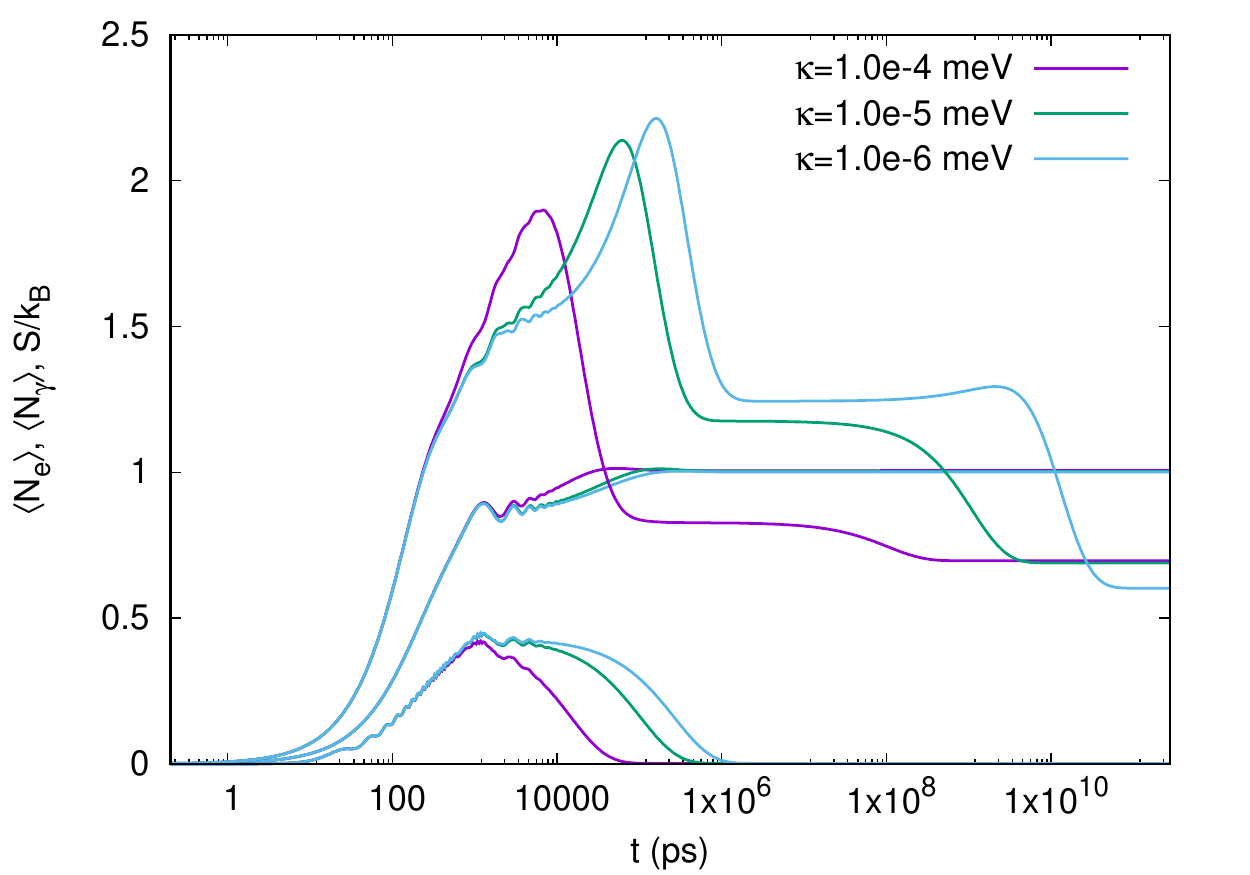}}
	\caption{The entropy $S/k_B$, the mean electron, and the mean photon number as functions of time for
	         the open central system for three different values of the cavity-environment coupling
             constant $\kappa$. For identification note, that in the steady-state the mean photon number seems to vanish, 
             and the mean electron approaches 1.  $g_\mathrm{EM} =0.1$ meV, $B=0.1$ T, $-eV_g=0.2$ meV, and $L_x=180$ nm.}
	\label{Fig05}
\end{figure}

The total mean number of electrons and photons is shown in Fig.\ \ref{Fig05} for three values of the 
cavity-environment coupling constant $\kappa$.
The mean electron number can be identified as represented by the curves approaching 1 in the steady-state
limit, but the curves representing the mean photon number seem to vanish in the same limit on the scale 
adopted for the figure. In addition, we show the R{\'e}niy-2 entropy of the central 
system \cite{2017arXiv170708946S,2018arXiv180608441B,2011arXiv1102.2098B}
\begin{equation}
      S=-k_B\ln{[Tr(\rho^2_\mathrm{S})]}.
\label{entropy}
\end{equation}
Interestingly, the entropy changes after the mean values of the total electron and photon numbers seem to have
reached their steady-state values. A closer inspection finds that there is a tiny change in the mean photon number
occurring in the same region as the last change in the entropy. 

\begin{figure}[htb]
	\centerline{\includegraphics[width=0.48\textwidth]{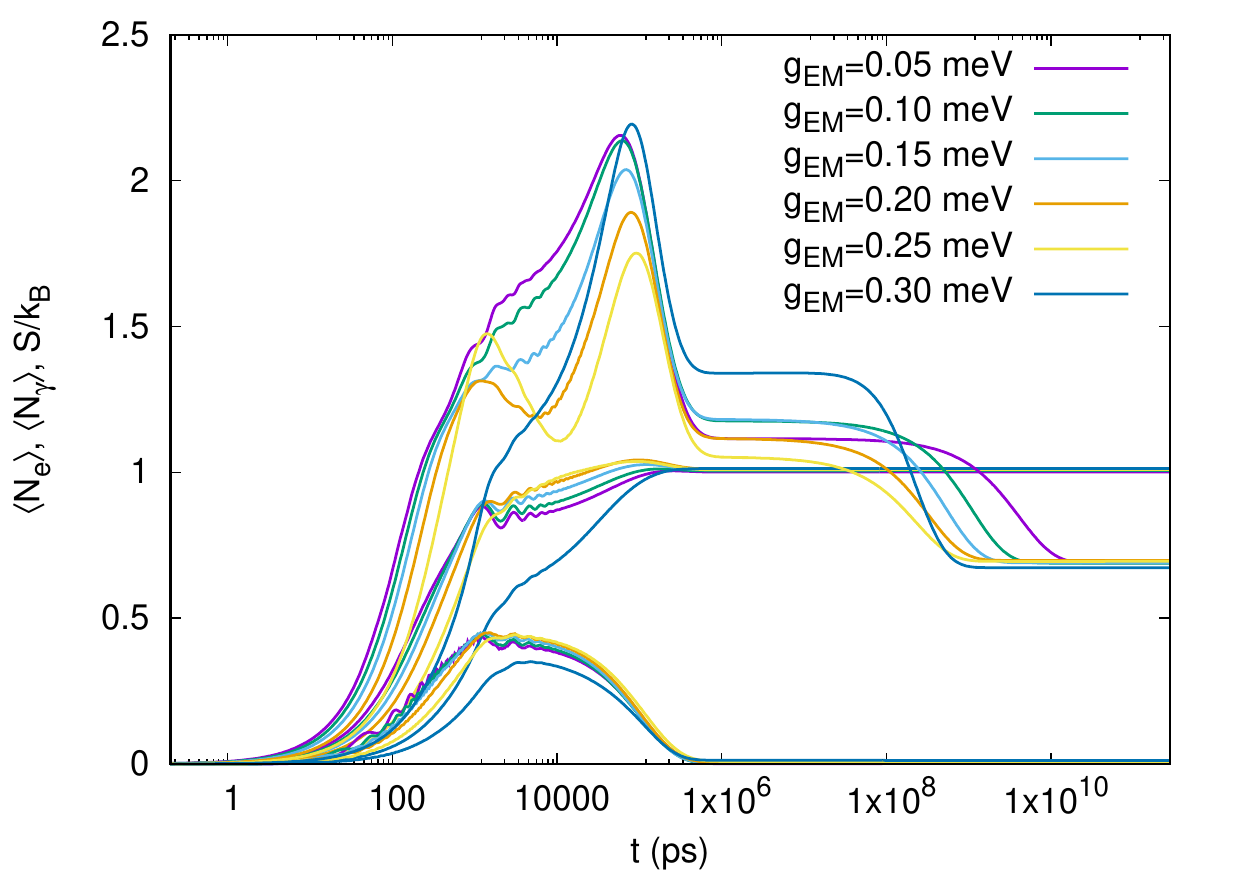}}
	\caption{The entropy $S/k_B$, the mean electron, and the mean photon number as functions of time for
		the open central system for six different values of the electron-photon coupling
		constant $g_\mathrm{EM}$. For identification note, that in the steady-state the mean photon number seems to vanish, 
		and the mean electron approaches 1. $\kappa =1\times 10^{-5}$ meV, $B=0.1$ T, $-eV_g=0.2$ meV, and $L_x=180$ nm.}
	\label{Fig06}
\end{figure}

For completeness, we display in Fig.\ \ref{Fig06} the same information as in Fig.\ \ref{Fig05}, but
here we vary the coupling coefficient $g_\mathrm{EM}$ and keep $\kappa =1\times 10^{-5}$ meV.
The conclusion from both Fig.\ \ref{Fig05} and \ref{Fig06} is that judged from the entropy the on-set of the steady-state
depends on both coupling parameters, $g_\mathrm{EM}$ and $\kappa$. This is not unexpected as they represent 
properties of the central system. The small oscillations visible in Fig.\ \ref{Fig05} and \ref{Fig06} in the range 
400 -- 10000 ps has been identified as coexisting nonequilibrium spin and Rabi oscillations \cite{Gudmundsson19:10}.  

For the record, we mention that in a system of two parallel quantum dots and initially with only one
photon found that after the mean electron and the mean current through the system seemed to have reached a steady-state
value an internal photon active transition changing the state of the system \cite{2016arXiv161109453G}.
In that case the system was not approaching the Coulomb blockage regime as the one-electron ground state was 
in the bias window, but initially there was no electron in the system, but one photon. 

\begin{figure}[htb]
	\centerline{\includegraphics[width=0.48\textwidth]{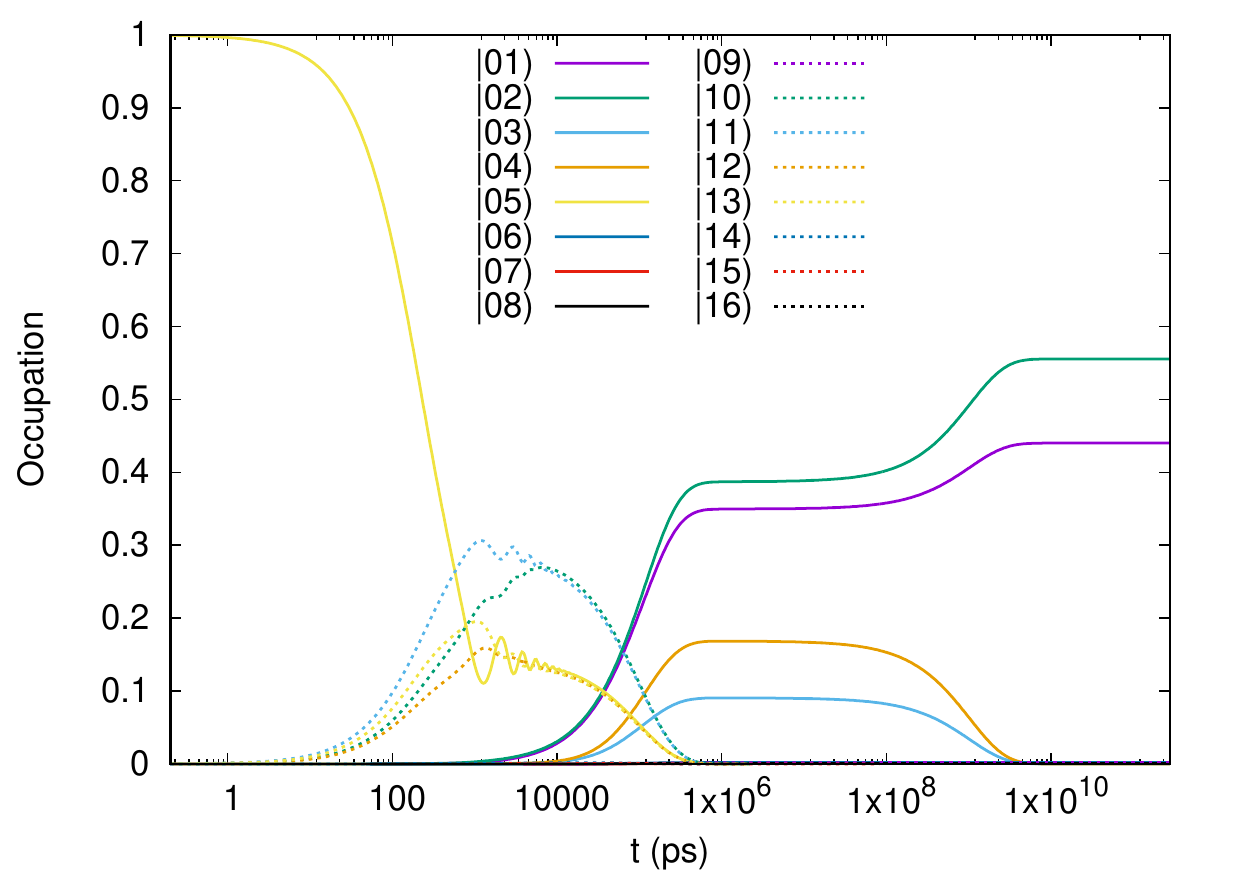}}
	\centerline{\includegraphics[width=0.48\textwidth]{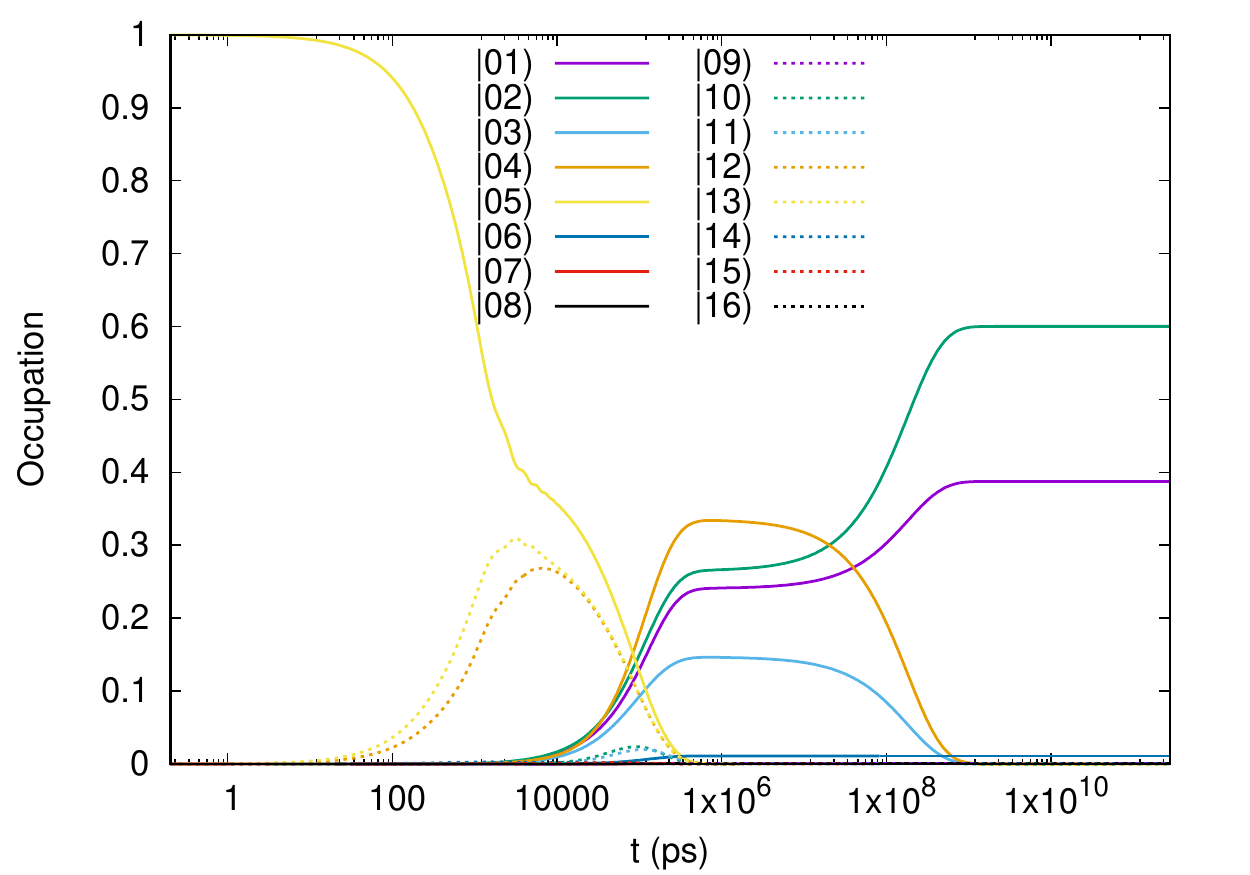}}
	\caption{The occupation of selected many-body states $|\breve{\mu})$ as a function of time
	         for $g_\mathrm{EM}=0.1$ meV (upper panel), and $g_\mathrm{EM}=0.3$ meV (lower panel).
             $\kappa =1\times 10^{-5}$ meV, $B=0.1$ T, $-eV_g=0.2$ meV, and $L_x=180$ nm.
             (The breve symbol is omitted here over the states).}
	\label{Fig07}
\end{figure}

The last slow transition bringing the system to the steady-state in the Coulomb blockage
regime can be identified by observing the time-dependent occupation of the many-body states
in Fig.\ \ref{Fig07}.
In the upper panel of Fig.\ \ref{Fig07} for $g_\mathrm{EM}=0.1$ meV we observe the empty state $|\breve{5})$
``loosing occupation'' as the states in the bias window, $|\breve{10})$, $|\breve{11})$, $|\breve{12})$, and 
$|\breve{13})$, gain occupation. These states in the bias window are Rabi-split with mean photon content close
to $1/2$ and feed the two spin components of the one-electron ground state, $|\breve{1})$, and $|\breve{2})$, mostly
localized in the right quantum dot, but also to a lesser extent the one-electron states, $|\breve{3})$, and 
$|\breve{4})$, mostly localized in the left quantum dot. The last transitions are thus from $|\breve{4})$ to
$|\breve{2})$, and from $|\breve{3})$ to $|\breve{1})$. For the higher electron-photon coupling 
$g_\mathrm{EM}=0.3$ meV, shown in the lower panel of Fig.\ \ref{Fig07}, similar processes take place with 
the exception that now an electron only enters into $|\breve{12})$, and $|\breve{13})$ as the states 
$|\breve{10})$ and $|\breve{11})$ are below the bias window. For this higher electron-photon coupling it
comes more likely than before for the electron to enter states $|\breve{3})$, and $|\breve{4})$, and the transition
to $|\breve{1})$, and $|\breve{2})$ goes faster. The lowering of states $|\breve{10})$ and $|\breve{11})$ 
below the bias window is the cause for the slower electron charging of the system in Fig.\ \ref{Fig06}
for the last curve at $g_\mathrm{EM}=0.30$ meV. 

\begin{figure}[htb]
	\centerline{\includegraphics[width=0.34\textwidth]{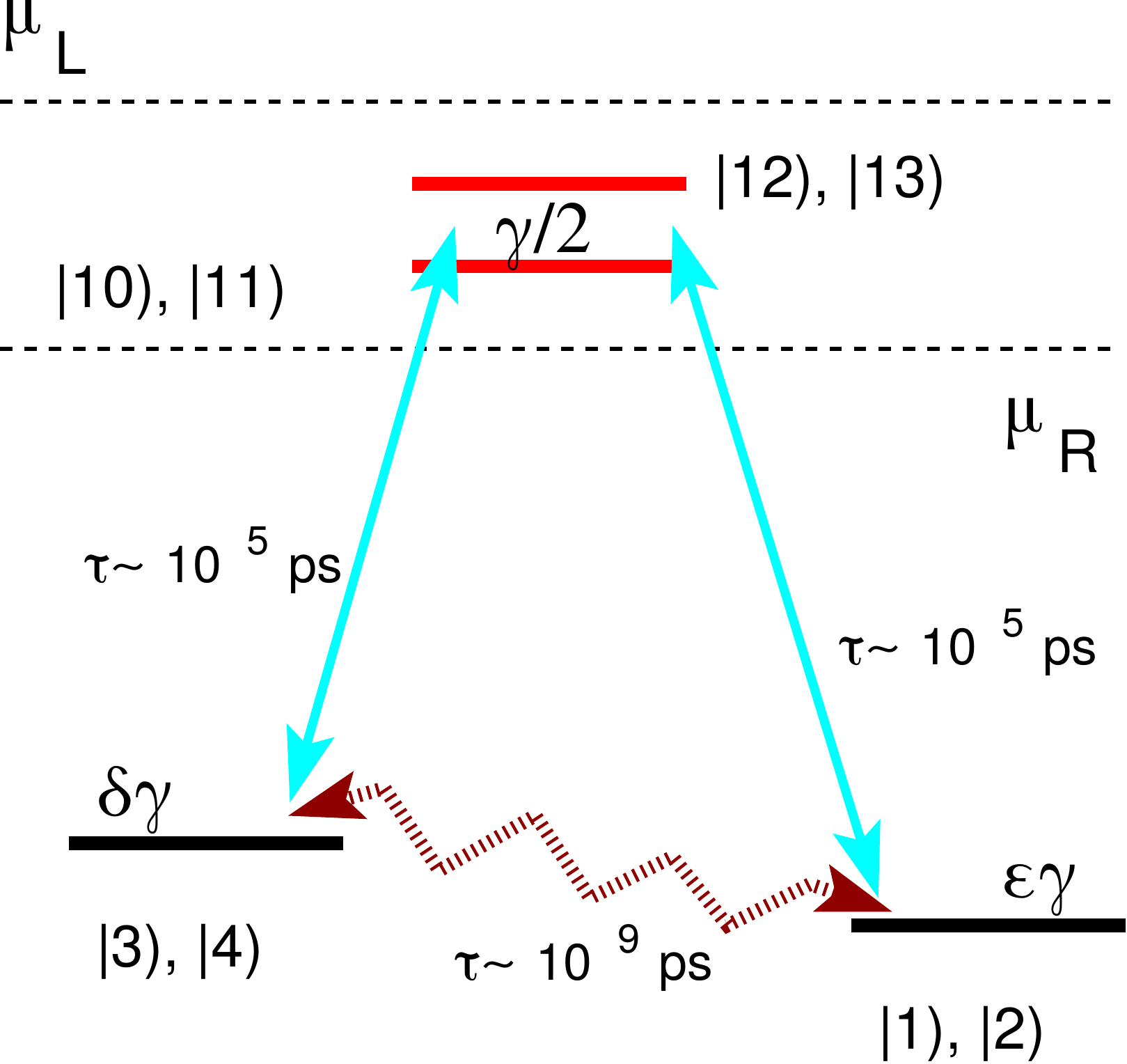}}
	\caption{Schema of the main transitions leading to the steady-state.
	         The azure arrows indicate the ``low order'' fast transitions connecting the 
	         Rabi-split states in the bias window to the lowest states in the quantum dots, 
	         but the wavy brown arrows indicate the 
	         high order slow transition connecting the lowest states mostly localized in
	         either the left or the right quantum dot. The approximate lifetime of the 
	         transitions ($\tau$) is indicated. (The breve symbol is omitted here over the states).}
	\label{Fig08}
\end{figure}

A schema of the main transitions relevant for the approach to the steady-state is shown
in Fig.\ \ref{Fig08}.
We now turn our attention to these slow transitions. For $g_\mathrm{EM}=0.05$ meV the mean photon number
for $|\breve{1})$, and $|\breve{2})$ is approximately $3.33\times 10^{-4}$ (indicated by $\epsilon\gamma$
in Fig.\ \ref{Fig08}), and for $|\breve{3})$, and $|\breve{4})$ the mean number is $3.68\times 10^{-4}$
(indicated by $\delta\gamma$ in Fig.\ \ref{Fig08}). Evaluated with a first order perturbation theory 
their mean photon number would vanish ($\epsilon =0$, $\delta =0$ in Fig.\ \ref{Fig08}), and no first 
order photon active transition would exist between them unless the cavity frequency is set at resonance 
$\omega=0.3413$ meV, see Eq.\ (\ref{H-e-EM-q}). 
The transitions to the 4 lowest one-electron states from the photon replica
states in the bias window do on the other hand exist in first order perturbation calculation.
Then the dynamics corresponding to the azure lines could be described by an effective three-level 
$\Lambda$ model \cite{Scully97} for each spin orientation. As these resonant transitions become 
inactive around $t\sim 1$ $\mu$s, the $\Lambda$-picture fails and the system enters an apparent 
steady-state in the time range $t\in [1,100]$ $\mu$s.
For the slow transitions, $|\breve{4})\rightarrow |\breve{2})$ and $|\breve{3})\rightarrow |\breve{1})$,
the exact paramagnetic matrix elements are approximately 5 orders of magnitude larger than the 
diamagnetic ones. The sensitivity of the lifetime of the upper states, $|\breve{3})$ and $|\breve{4})$,
on the electron-photon coupling makes us conclude that only some photon active higher order transitions are
possible with a stronger para- than diamagnetic character.
Supporting this claim is the behavior of the lifetime when the cavity-environment coupling $\kappa$
is varied (see Fig.\ \ref{Fig05}). The shortened lifetime with increasing $\kappa$ reflects the 
Purcell effect on photon active transitions \cite{PhysRev.69.681}, which was recently predicted to be visible
in the transport current as a function of the photon energy in the steady-state \cite{Nzar-2019-Rabi}.
The influence of $\kappa$ on the slow transition is clearly seen in the eigenspectrum of the 
Liouvillian in Fig.\ \ref{Fig04}.

\begin{figure}[htb]
	\centerline{\includegraphics[width=0.48\textwidth]{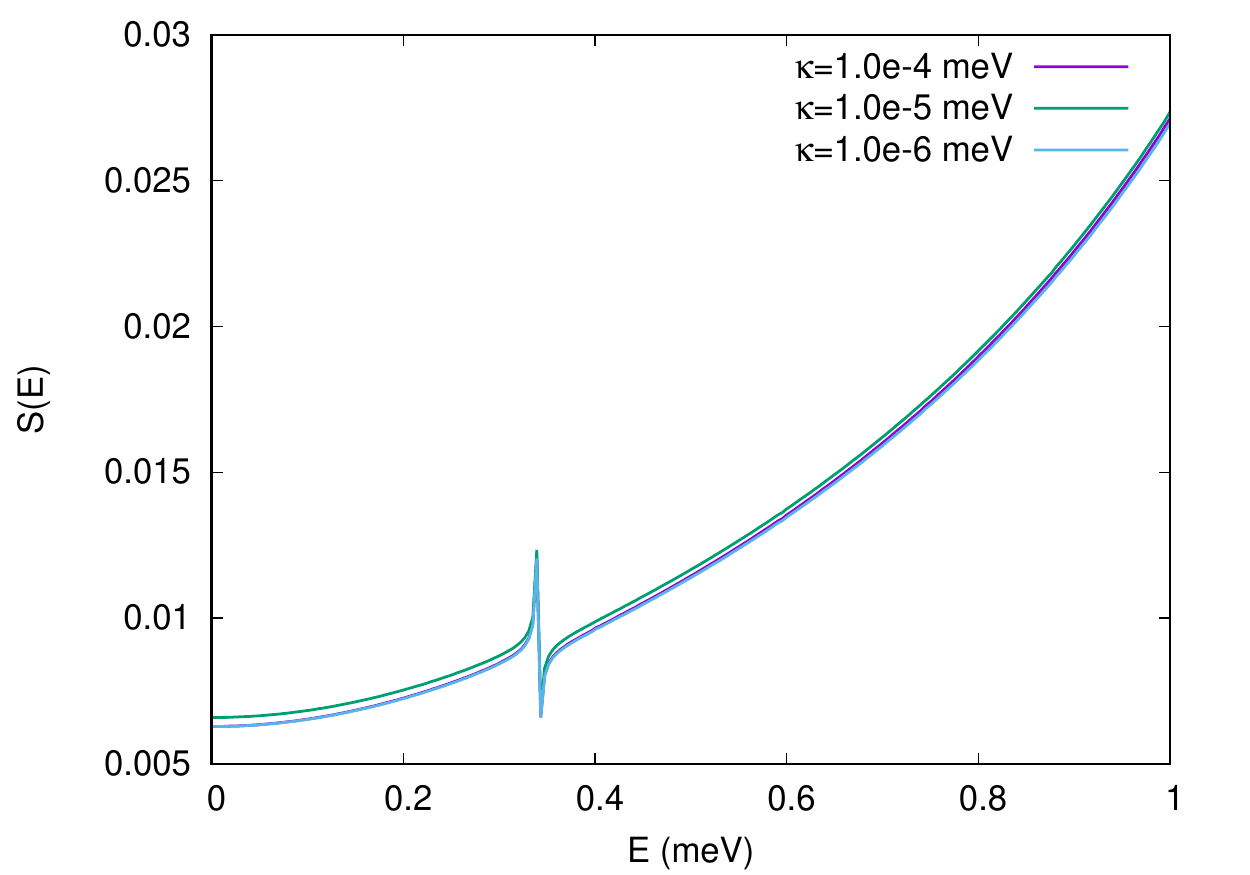}}
	\centerline{\includegraphics[width=0.48\textwidth]{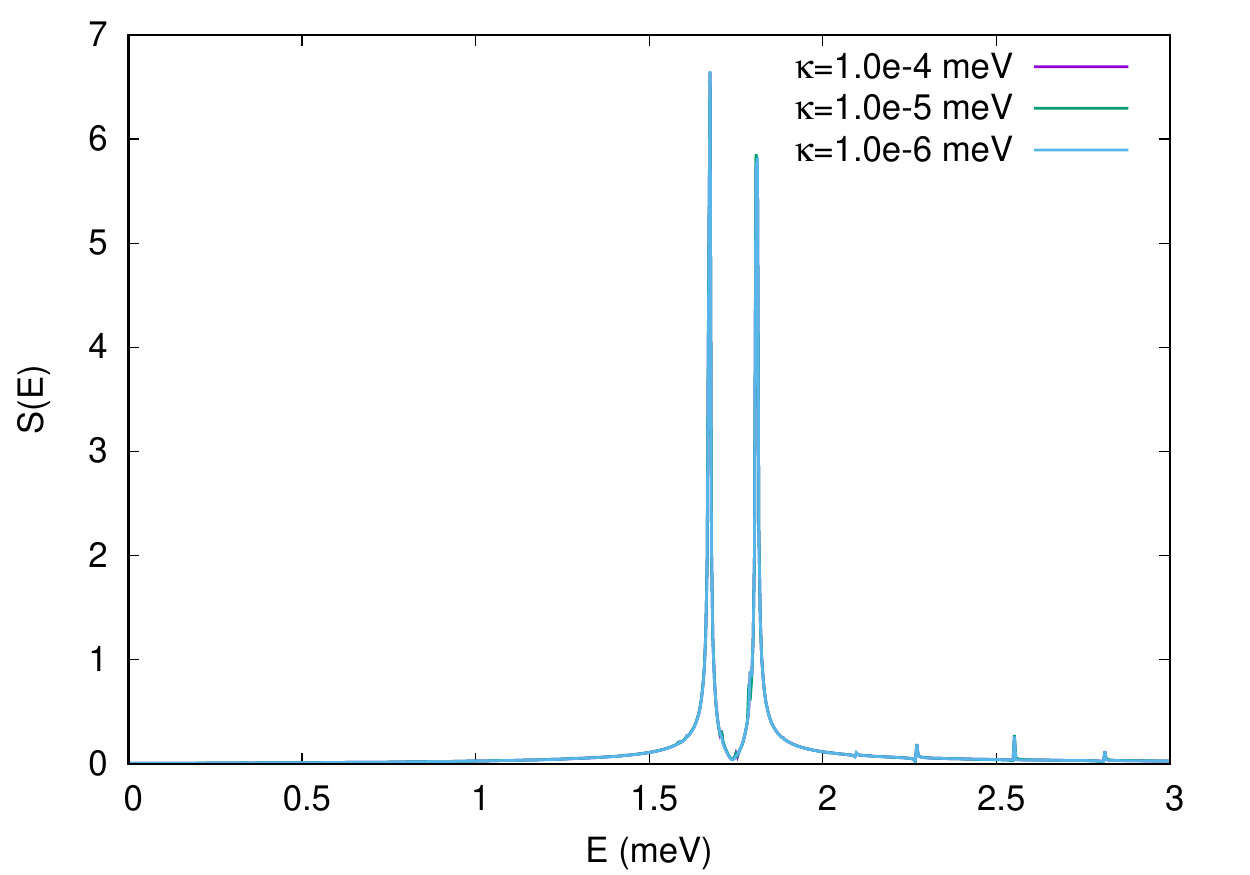}}
	\caption{The Fourier power spectra for the photon correlation function
	         $S(E)$ in the steady state for three values of the cavity-environment coupling
             $\kappa$. $g_\mathrm{EM}=0.1$ meV, $B=0.1$ T, $-eV_g=0.2$ meV, and $L_x=180$ nm.}
	\label{Fig09}
\end{figure}

In the steady-state the Fourier spectral density of the emitted cavity radiation \cite{doi:10.1002/andp.201700334}
\begin{equation}
      S(E) = \frac{\kappa}{\pi}\left|\int_0^\infty \frac{d\tau}{\hbar}e^{-iE\tau /\hbar}
             \{\langle X (\tau) X(0)\rangle\}\right| , 
\label{SE}
\end{equation}   
with $X=a+a^\dagger$, is displayed in Fig.\ \ref{Fig09}.
On the coarser scale used in the lower panel of Fig.\ \ref{Fig09} are seen the two peaks
split by approximately 0.135 meV, but centered around the energy of the cavity mode $\hbar\omega = 1.75$ meV.
The splitting corresponds to the Rabi splitting read from the energy spectrum. 
The finer scale used in the upper panel reveals a Fano-like resonance peak at the energy 0.3413 meV 
corresponding to the energy of
the last slow transitions $|\breve{4})\rightarrow |\breve{2})$ and $|\breve{3})\rightarrow |\breve{1})$.
The interpretation is that a very weak electromagnetic perturbation of the system in the steady-state
(consisting mainly of the one-electron ground state) activates the 
transitions to the states in the bias window and the very slow transition from the spin components of
the one-electron ground state to the states $|\breve{3})$ and $|\breve{4})$. Both types of excitations
are radiative, or photon active, but are of a very different strength.

\begin{figure}[htb]
	\centerline{\includegraphics[width=0.48\textwidth]{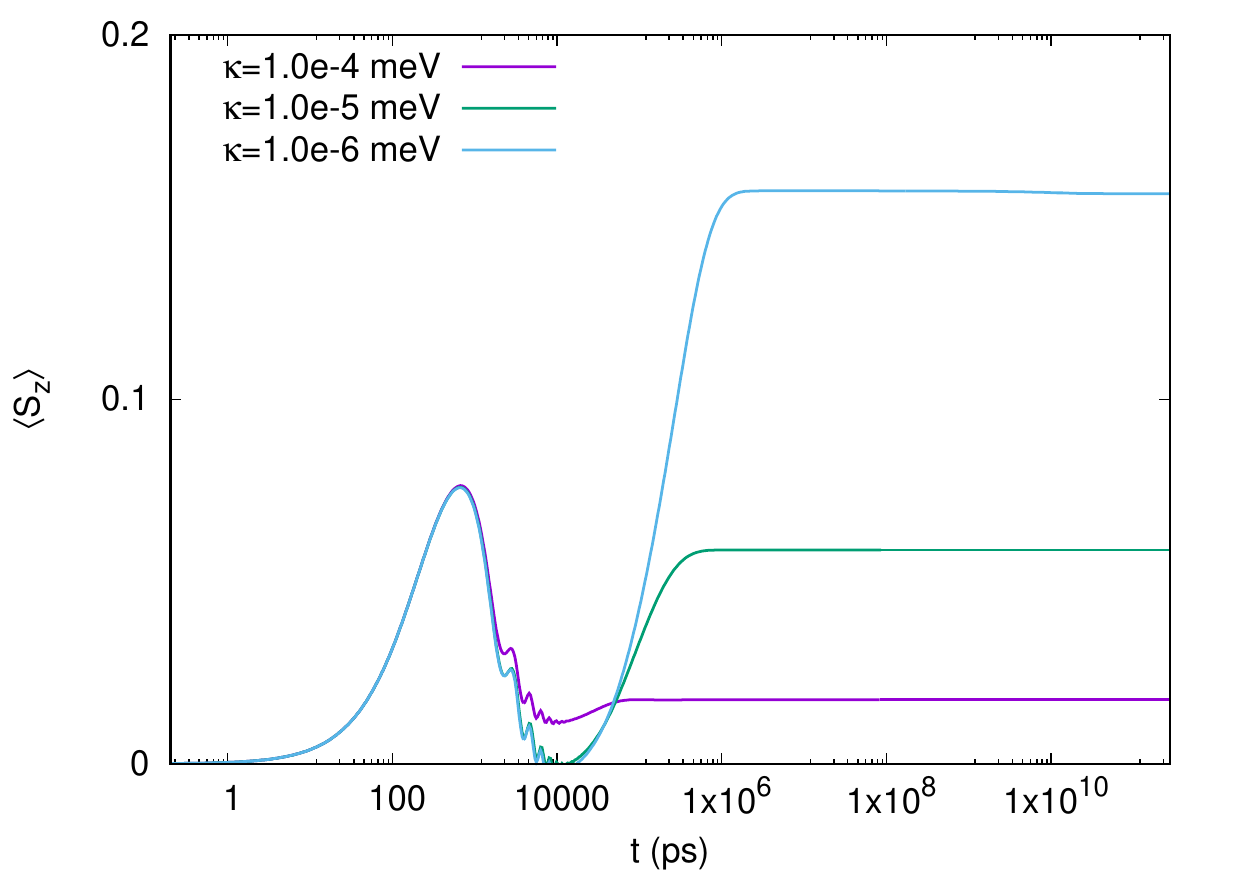}}
	\caption{The mean $z$-component of the total spin, $S_z/\hbar$, as function of time
		     for three different values of the cavity-environment coupling $\kappa$. 
	         $g_\mathrm{EM}=0.1$ meV, $B=0.1$ T, $-eV_g=0.2$ meV, and $L_x=180$ nm.}
	\label{Fig10}
\end{figure}

Last, a note on a curious emergence of a spin-polarization that can be seen in the steady-state
in Fig.\ \ref{Fig07}, where the polarization is increased in the lower panel, i.e.\ for a larger
electron-photon coupling $g_\mathrm{EM}$. In Fig.\ \ref{Fig10} is plotted the mean value of the 
$z$-component of the spin as a function of time for three values of the cavity-environment 
coupling $\kappa$.
Clearly, the spin polarization in Fig.\ \ref{Fig07} is not caused by thermal effects,
instead it is important to have in mind that the external semi-infinite leads are 
defined by a potential and have thus a quasi-one-dimensional electron system with
subband structure. In addition, the system is in a weak external magnetic field
$B=0.1$ T, and the probability of tunneling into the states in the bias window
depends on the density of states of the leads and the shape of the states in the 
central electronic system. The small increase of the spin-polarization (note the 
small vertical scale of Fig.\ \ref{Fig10}) is thus influenced by many factors 
as the path to the steady-state differs for different values of $\kappa$.
This behavior can be related to the nonequilibrium spin-oscillations discussed
earlier \cite{Gudmundsson19:10}.

\section{Summary}
\label{Sec4}
In this report we have shown that in order to describe the time-evolution of electron
transport through a nanoscale system in a GaAs heterostructure embedded
in a terahertz cavity it is necessary to treat the electron-photon interaction 
nonperturbatively to make sure none of the vital transition is missed that
takes the system finally to its steady state. On the long timescale needed some
transitions forbidden in low order perturbation can be essential. Though the 
transition we take as an example has different symmetry, the difference in timescales
possible are well know from, for example, the 2S$\rightarrow$1S in atomic 
hydrogen \cite{PhysRev.116.363,Choi_1987,PhysRevLett.107.203001}.

As expected in the terahertz range, the exact matrix elements for a rectangular
photon-cavity do not change our results much, compared to matrix elements that
are calculated assuming the field to be constant over the sample size of the 
nanoscale system. This is true as we anyway in both cases use a numerically
exact diagonalization scheme to calculate the cavity-photon dressed many-body 
electron states and their energies. The higher order contributions are more
important than the much smaller corrections caused by the shape of the cavity.
But, it is important to remember that the exact matrix elements for the diamagnetic
are not only diagonal in the electronic variables, and thus it is possible to find a 
transition caused by this term, that would otherwise go unnoticed, and might be comparable to the 
2S$\rightarrow$1S transition in atomic hydrogen. Even, in the approximation that
the electronic part of the interaction is diagonal the total interaction is not 
and can thus cause a tiny Rabi-splitting if that can be seen if the paramagnetic
interaction is blocked by symmetry \cite{doi:10.1002/andp.201700334}.

We interprete the results from our continuous model as showing a high
order photon-cactive transtion between the ground states of weakly coupled 
quantum dots.
In addition, we reconfirm the effects of the Purcell effect in a transport 
current \cite{Nzar-2019-Rabi}, and discover a slight spin polarization
dependent on the cavity-decay or cavity-environment coupling constant
$\kappa$ connected to non-equilibrium oscillations of the spin reported
elsewhere \cite{Gudmundsson19:10}. We emphasize that even when we talk about
photon active transitions it is important to remmeber that the coupling to 
the leads is a neccesary trigger in order to perturb the exact fully interacting
eigenstates of the central system.

We hope to convey the message to the reader that details in the many-body
description of the electron-photon interaction together with the geometry
or shape of the systems can be of importance to understand promising
electron-photonic systems, and through them we can hope to obtain better 
fundamental understanding of these interactions.

\begin{acknowledgments}
This work was financially supported by the Research Fund of the University of Iceland,
the Icelandic Research Fund, grant no.\ 163082-051, 
and the Icelandic Instruments Fund. The computations were performed on 
resources provided by the Icelandic High Performance Computing Centre at the University of Iceland.
V.M.\ acknowledges financial support from CNCS - UEFISCDI grant PN-III-P4-ID-PCE-2016-0084, and from 
the Romanian Core Program PN19-03 (contract no.\ 21 N/08.02.2019)
\end{acknowledgments}

\appendix
\section{Closed system: Technical details}
\label{AppA}
The integrals for the matrix elements of the electron-photon interactions 
have integrands of three trigonometric functions
over a finite interval or two Hermite polynomials, a Gaussian exponent and a trigonometric function 
over the infinite $y$-axis. After the matrix elements are found they are first transformed unitairily 
to the basis of 2688 exact single-electron functions for the short quantum wire in a perpendicular 
homogeneous magnetic field. Subsequently, after the construction of the 683 dimensional Fock many-body 
space using the 36 lowest in energy single-electron states, and the diagonalization of the Coulomb interaction 
therein the electron-photon matrix elements are transformed unitairily to the 512 dimensional
truncated Fock space of exactly Coulomb interacting electrons. This last Fock space includes states
with one, two and three electrons in a ratio adequate to its energy extension. The exact Coulomb
interacting Fock space is then tensor multiplied by the 17 lowest eigenstates of the photon operator
to form the Fock space used to diagonalize the electron-photon interaction in resulting in a
8704 dimensional many-body space of cavity-photon dressed electrons, of which we will use the 128
lowest in energy states to perform the transport calculations in. 
This step wise construction of the relevant many-body spaces and their truncation is necessary as
a single-step construction would have required a much too large basis to attain an acceptable
convergence, and an important fact to keep in mind is that the construction is much more sensitive 
to the number of electron states kept in the calculation than photon states, whose number is simple
to increase. The reason for this is the polarization of charge by the electron-photon interaction.
A rotating wave approximation is not used for the electron-photon interaction as in a complex
central system with many states there might always be some transitions close to a Rabi resonance
and other far from it.

\begin{figure}[htb]
	{\includegraphics[width=0.23\textwidth]{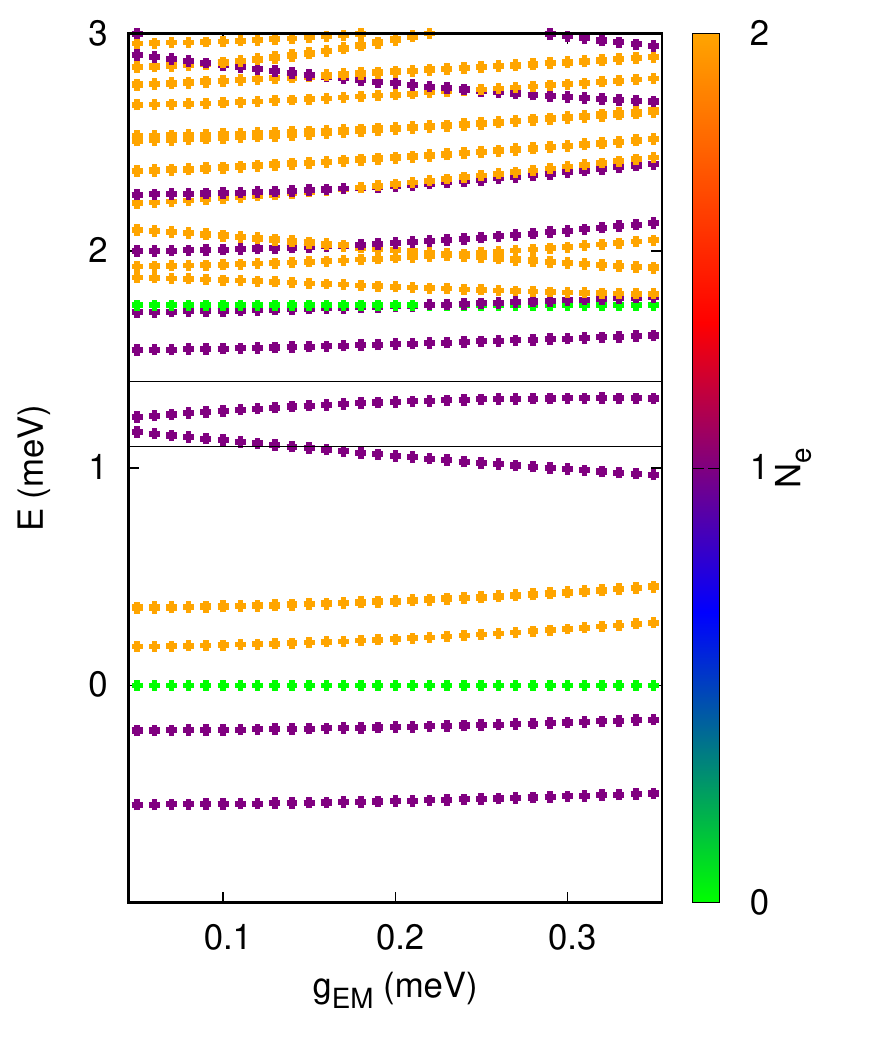}}
	{\includegraphics[width=0.23\textwidth]{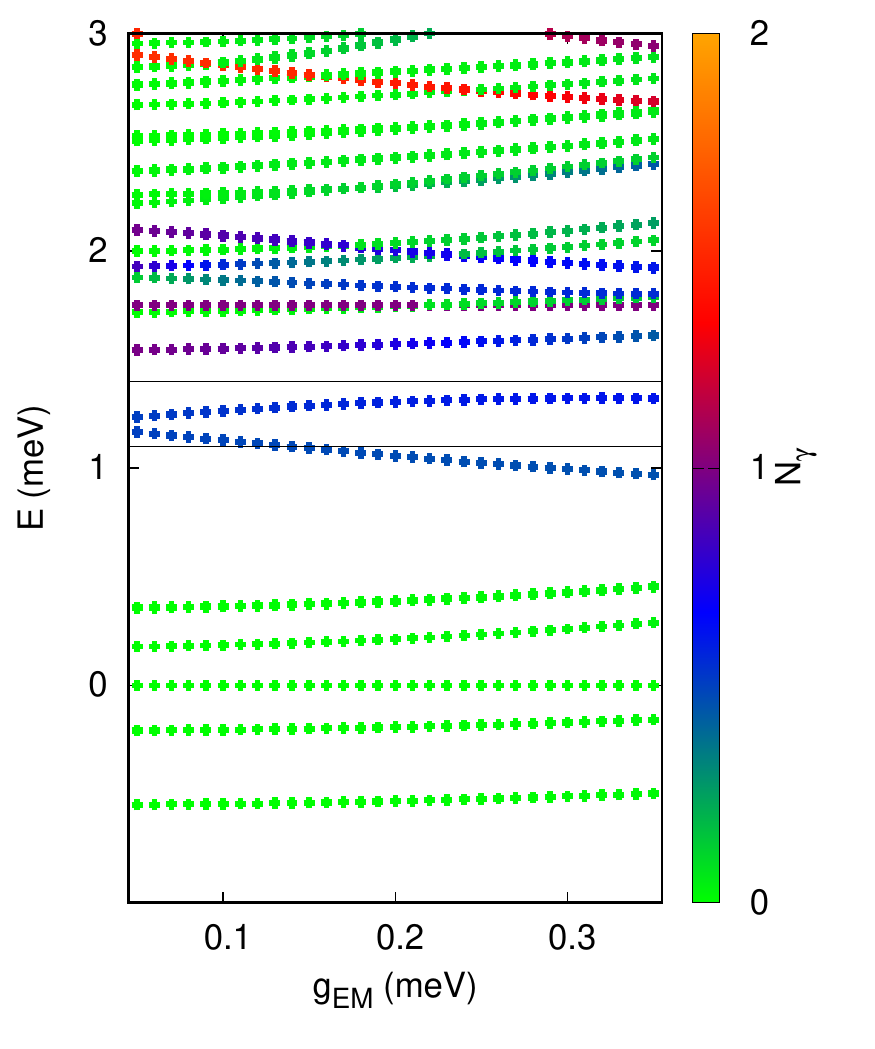}}
	{\includegraphics[width=0.23\textwidth]{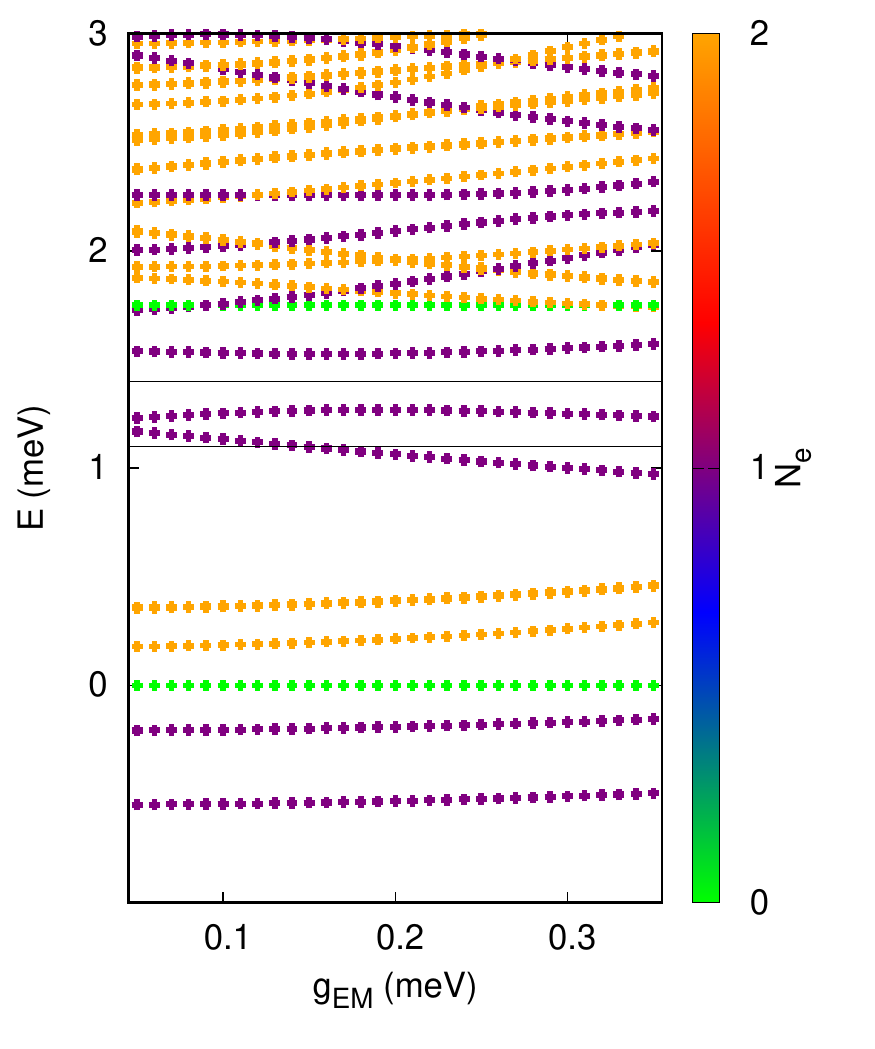}}
	{\includegraphics[width=0.23\textwidth]{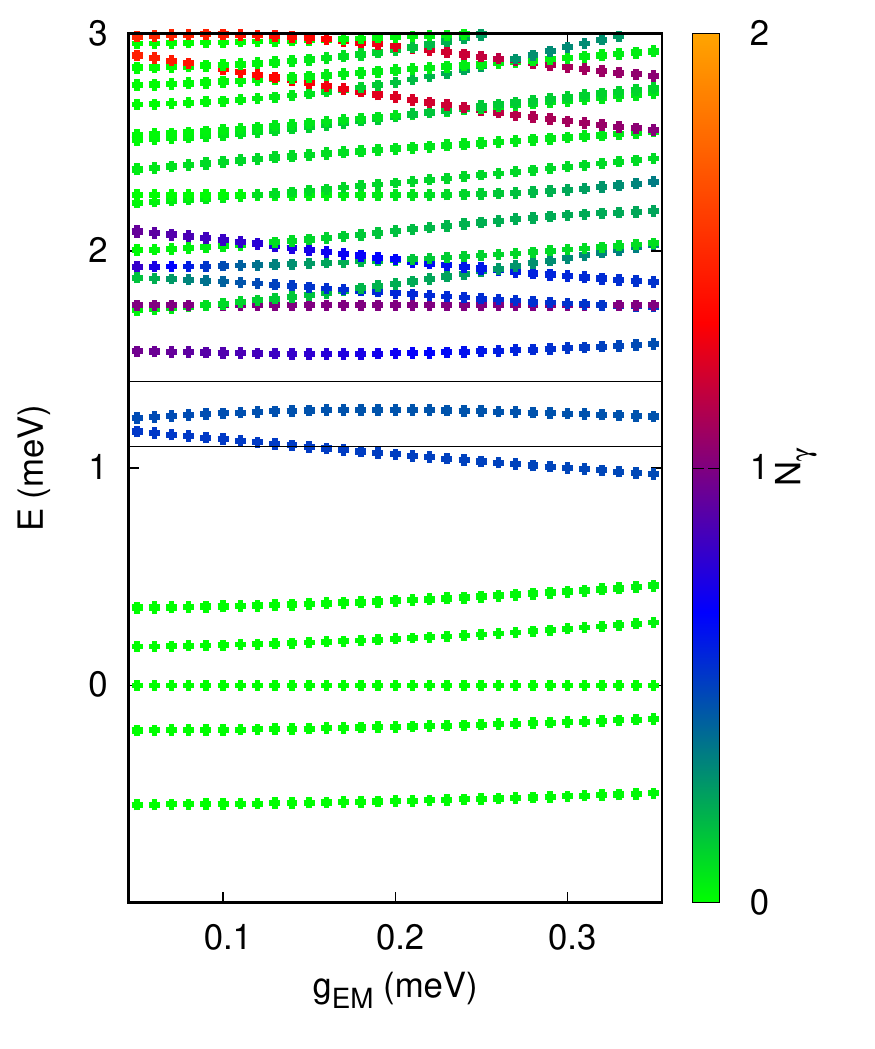}}
	\caption{The many-body energy spectra as functions of the electron-photon coupling strength $g_\gamma$
	         for $x$-polarized (upper panels), and $y$-polarized (lower panels) cavity field. The mean electron number
             is color coded in the spectra (left panels), and the mean photon number (right panels). For a closed system
             the electron number is an integer, but not the photon number. $B=0.1$ T, $L_x=180$ nm, 
             $\hbar\Omega_0=2.0$ meV, $-eV_g=0.2$ meV, and $\hbar\omega =1.75$ meV.}
	\label{Fig11}
\end{figure}

The energy spectra as functions of the electron-photon coupling parameter $g_\mathrm{EM}$ for the 
many-body states of the fully interacting system are displayed in
Fig.\ \ref{Fig11} for $x$-polarization of the cavity-photon field in the upper panels and 
$y$-polarization in the lower panels.
The many-body states $|\breve{\mu})$ are cavity-photon dressed electron states with an integer number of
electrons indicated in the left panels, but their average photon content is shown in the right panels.

\section{Transport: Technical details}
\label{AppB}
The time evolution the couplings to the reservoirs induce are best described by the Liouville-von 
Neumann equation for the probability operator (the density operator) of the total system,
but as the continuous character of the respective reservoirs make the resulting Fock-space much
too large we resort to the projection formalism of Nakajima \cite{Nakajima58:948} 
and Zwanzig \cite{Zwanzig60:1338}. Initially, we derive a non-Markovian generalized master equation (GME) for
the reduced density operator of the central system including terms up to second order in the 
lead-system coupling (\ref{H_T}) in the kernel of the resulting integro-differential 
equation \cite{Gudmundsson12:1109.4728}. Subsequently, we apply vectorization \cite{IMM2012-03274} and Kronecker
tensor products together with a Markovian approximation to transform the GME from the Fock-space
of many-body states to the Liouville space of transitions \cite{2016arXiv161003223J,Weidlich71:325}. 
We include 128 Fock-states in our transport calculations and end thus up with 16384 transitions
in the Liouville-space. The increased space size is counteracted by efficient parallelization 
and GPU-processing \cite{2016arXiv161003223J}. The weak photon dissipation is derived with a
Markovian and rotating wave approximation, where the creation(annihilation) operator for the 
cavity-photons needs to be rid of fast rotating annihilation(creation) terms when transformed
to the fully interacting basis $|\breve{\mu})$ in order for vacuum processes to be correctly 
described in the model \cite{PhysRevA.84.043832,PhysRevA.31.3761,PhysRev.129.2342,Gudmundsson19:10}. 

Due to the vectorization \cite{IMM2012-03274} the Markovian master equation in Liouville-space assumes 
the form of a simple linear first order differential equation \cite{2016arXiv161003223J,PhysRevB.81.155303}
\begin{equation}
    \partial_t \vctrz(\rho_\mathrm{S}) = -i\mathfrak{L} \vctrz(\rho_\mathrm{S}),
\label{finalME}
\end{equation}
which has an analytical solution that is convenient to use to search for the steady-state of the system.
\begin{equation}
    \vctrz(\rho_\mathrm{S}(t)) =
    \left\{\mathfrak{U}\left[\exp{\left(-i\mathfrak{L}_\text{diag}t\right)}\right]\mathfrak{V}\right\}
    \vctrz(\rho_\mathrm{S}(0)),
\label{eq:solution}
\end{equation}
with the left and right eigenvector matrices of the nonhermitian Liouville operator satisfying 
$\mathfrak{UV}=I$ and $\mathfrak{VU}=I$ \cite{2016arXiv161003223J}. The imaginary part of the 
eigenspectrum of the Liouville operator $\mathfrak{L}$ reveals the 16384 relaxation coefficients
at work in the Markovian time evolution of the system.

%
%
\frenchspacing
\bibliographystyle{apsrev4-1}
%

%
%
%
\end{document}